\documentstyle[12pt]{article}
\def \Et {{\rm E}_{\rm T}}

\def\Z0{${\em Z^0\/}$}

\def\r#1 {$^{#1}$}

\newcommand{\et}{{\rm E}_{\scriptscriptstyle\rm T}}

\newcommand{\met}{\mbox{$\protect \raisebox{.3ex}{$\not$}\et$}}

\newcommand{\ppbar}{p\bar{p}}

\newcommand{\ttbar}{t\bar{t}}
\newcommand{\bbbar}{b\bar{b}}

\newcommand{\geapp}{{\rm\raisebox{-0.7ex}{$\stackrel{\textstyle >}{\sim}$}}}

\begin{document}
\begin{flushright}
FERMILAB-PUB-94/116-E \\
CDF/PUB/TOP/PUBLIC/2595 \\
May 16, 1994 \\
\end{flushright}

\begin{center}
{\large Evidence for Top Quark Production in $\bar{p} p$ Collisions 
at $\sqrt{s}=1.8$~TeV}         
\end{center}
\vspace{0.5cm}

\font\eightit=cmti8
\def\r#1{\ignorespaces $^{#1}$}
\hfilneg
\begin{sloppypar}
\noindent
F.~Abe,\r {13} M.~G.~Albrow,\r 7 S.~R.~Amendolia,\r {23} D.~Amidei,\r {16} 
J.~Antos,\r {28}
C.~Anway-Wiese,\r 4 G.~Apollinari,\r {26} H.~Areti,\r 7 
P.~Auchincloss,\r {25} M.~Austern,\r {14}
% ...
\end{sloppypar}

\vskip .025in
\begin{center}
(CDF Collaboration)
\end{center}

\vskip .025in
\begin{center}
\r 1  {\eightit Argonne National Laboratory, Argonne, Illinois 60439} \\
\r 2  {\eightit Brandeis University, Waltham, Massachusetts 02254} \\
\r 3  {\eightit Istituto Nazionale di Fisica Nucleare, University of Bologna,
I-40126 Bologna, Italy} \\
\r 4  {\eightit University of California at Los Angeles, Los 
Angeles, California  90024} \\  
\r 5  {\eightit University of Chicago, Chicago, Illinois 60637} \\
\r 6  {\eightit Duke University, Durham, North Carolina  27708} \\
\r 7  {\eightit Fermi National Accelerator Laboratory, Batavia, Illinois 
60510} \\
\r 8  {\eightit Laboratori Nazionali di Frascati, Istituto Nazionale di Fisica
               Nucleare, I-00044 Frascati, Italy} \\
\r 9  {\eightit Harvard University, Cambridge, Massachusetts 02138} \\
\r {10} {\eightit University of Illinois, Urbana, Illinois 61801} \\
\r {11} {\eightit Institute of Particle Physics, McGill University, Montreal 
H3A 2T8, and University of Toronto,\\ Toronto M5S 1A7, Canada} \\
\r {12} {\eightit The Johns Hopkins University, Baltimore, Maryland 21218} \\
\r {13} {\eightit National Laboratory for High Energy Physics (KEK), Tsukuba, 
Ibaraki 305, Japan} \\
\r {14} {\eightit Lawrence Berkeley Laboratory, Berkeley, California 94720} \\
\r {15} {\eightit Massachusetts Institute of Technology, Cambridge,
Massachusetts  02139} \\   
\r {16} {\eightit University of Michigan, Ann Arbor, Michigan 48109} \\
\r {17} {\eightit Michigan State University, East Lansing, Michigan  48824} \\
\r {18} {\eightit University of New Mexico, Albuquerque, New Mexico 87131} \\
\r {19} {\eightit Osaka City University, Osaka 588, Japan} \\
\r {20} {\eightit Universita di Padova, Instituto Nazionale di Fisica 
          Nucleare, Sezione di Padova, I-35131 Padova, Italy} \\
\r {21} {\eightit University of Pennsylvania, Philadelphia, 
        Pennsylvania 19104} \\   
\r {22} {\eightit University of Pittsburgh, Pittsburgh, Pennsylvania 15260} \\
\r {23} {\eightit Istituto Nazionale di Fisica Nucleare, University and Scuola
               Normale Superiore of Pisa, I-56100 Pisa, Italy} \\
\r {24} {\eightit Purdue University, West Lafayette, Indiana 47907} \\
\r {25} {\eightit University of Rochester, Rochester, New York 14627} \\
\r {26} {\eightit Rockefeller University, New York, New York 10021} \\
\r {27} {\eightit Rutgers University, Piscataway, New Jersey 08854} \\
\r {28} {\eightit Academia Sinica, Taiwan 11529, Republic of China} \\
\r {29} {\eightit Superconducting Super Collider Laboratory, Dallas, 
Texas 75237} \\
\r {30} {\eightit Texas A\&M University, College Station, Texas 77843} \\
\r {31} {\eightit University of Tsukuba, Tsukuba, Ibaraki 305, Japan} \\
\r {32} {\eightit Tufts University, Medford, Massachusetts 02155} \\
\r {33} {\eightit University of Wisconsin, Madison, Wisconsin 53706} \\
\r {34} {\eightit Yale University, New Haven, Connecticut 06511} \\
\end{center}

\begin{abstract}
We summarize a search\cite{prd} for the top quark with the Collider Detector 
at Fermilab (CDF) in a sample of 
$\bar{p}p$ collisions at $\sqrt{s}$= 1.8 TeV with an integrated
luminosity of 19.3~pb$^{-1}$.
We find 12 events consistent with 
either two $W$ bosons, or a $W$ boson and at least one $b$ jet. 
The probability that the measured yield is
consistent with the background is 0.26\%.
Though the statistics are too limited to establish firmly the existence
of the top quark, a natural interpretation of the excess
is that it is due to $t\bar{t}$ production. Under this assumption,
constrained fits to individual events yield a 
top quark mass of $174 \pm 10^{+13}_{-12}$ GeV/c$^2$. 
The $t\bar{t}$ production cross section is measured to be 
$13.9^{+6.1}_{-4.8}$~pb. 

\end{abstract}



The Standard Model has enjoyed outstanding success, yet 
the top quark, which is required as the weak-isospin
partner of the bottom quark, has remained unobserved.  
Direct searches at the Fermilab Tevatron have placed a 95\%
confidence level lower limit of $M_{top}>131$~GeV/c$^2$\cite{d0_limit}.
Global fits to precision electroweak measurements
yield a favored mass of 
\mbox{$M_{top} = 174 ^{+11 +17}_{-12 -19}$~GeV/c$^2$}~\cite{Moriond}.

One expects that, at Tevatron energies, 
most top quarks are produced in pairs.
For $M_{top}~\geapp~85$~GeV/c$^2$, each top quark decays 
to a real $W$ boson and a $b$ quark. The observed event topology
is then determined by the decay mode of the two $W$ bosons.
About 5\% of the time both $W$ bosons decay to $e\nu$
or $\mu\nu$ (the ``dilepton mode"), giving two 
high-$P_T$ leptons with opposite charge, two $b$ jets, and large
missing transverse energy ($\met$)
from the undetected neutrinos\cite{coord}.
In another 30\% of the cases one
$W$ boson decays to $e\nu$ or $\mu\nu$, and the other 
to a $q\bar{q}^{\prime}$ pair (the ``lepton+jets mode"). 
This final state includes a high-$P_T$ charged lepton, $\met$, and
jets from the $W$ and the two $b$~quarks. 
The remaining 65\% of the final states involve the hadronic 
decays of both
$W$ bosons, or the decay of one or both of the $W$ bosons into $\tau$ leptons.
These channels have larger backgrounds
and are not considered here. This analysis is based on 
a sample of $\bar{p}p$ collisions at $\sqrt{s}=1.8$~TeV with an 
integrated luminosity of 19.3$\pm$0.7~pb$^{-1}$,
collected at the Fermilab 
Tevatron by the CDF detector\cite{cdf_det} in 1992-3. The details
of the analysis are presented in Ref.~\cite{prd}.

The momenta of charged particles are measured in the central tracking
chamber (CTC), which sits inside a 1.4-T superconducting solenoidal
magnet. Outside the CTC, electromagnetic and hadronic calorimeters,
arranged in a projective tower geometry, cover the pseudorapidity
region $|\eta|<3.6$, allowing
reliable measurements of the $\met$. The calorimeters are
also used to identify jets and electron 
candidates. Outside the calorimeters, drift chambers in the region
$|\eta|<1.0$ provide muon identification. 
A silicon vertex detector (SVX)\cite{SVX_paper}, located immediately
outside the beampipe, provides
precise track reconstruction in the plane transverse to the beam,
and is used to identify 
secondary vertices that can be produced by
$b$ and $c$ quark decays.  A three-level trigger selects the 
inclusive electron and muon events used in this analysis.

%

In the dilepton search, both leptons are required to have 
$P_T > 20$~GeV/c and to have opposite charge.
At least one of the leptons is required to have
$|\eta|<1.0$ and to be isolated\cite{prd}.
In addition, we require $\met>25$~GeV\cite{jetcor}.
To remove background from $Z$ production, we reject 
$ee$ and $\mu\mu$ events with  
\mbox{75 $ < M_{\ell\,\ell} <  $  105 GeV/c$^2$}. For $M_{top}>120$~GeV/c$^2$,
the two $b$ quarks have significant energy and are 
detected with good efficiency as jets. By requiring 
two jets with $|\eta| < 2.4$ and $E_{T} > 10 $ GeV\cite{jetcor},
we reduce backgrounds by a 
factor of four while preserving 84\% of the signal for
$M_{top}=160$~GeV/c$^2$.
To achieve additional rejection against
$Z\rightarrow\tau\tau$ events and events with $\met$ 
induced by jet mismeasurement,
we require, for $\met < $ 50~GeV, that the azimuthal angle 
between the $\met$ and the nearest lepton or jet exceed 20$^\circ$. 
No $ee$ or $\mu\mu$ events pass all cuts. Two $e\mu$ events survive.

We use the ISAJET\cite{Paige} Monte Carlo program to determine the 
acceptance and the efficiency of the event-selection criteria.
The fractional uncertainty in the efficiency of the two-jet requirement, 
due mostly to the limited understanding of gluon radiation, decreases
from 13$\%$ for $M_{top}$=120 GeV/c$^2$ to 3$\%$ for $M_{top}$=180 GeV/c$^2$.
Other uncertainties in the detection efficiency 
come from the lepton-identification cuts (6\%),
lepton-isolation cuts (2\%), $\met$ cuts (2\%), 
structure functions (2\%), and 
Monte Carlo statistics (3\%).
The overall acceptance, $\epsilon_{DIL}$, for the dilepton 
search is shown in Table~\ref{sum_acc}. The number of expected dilepton
events from $\ttbar$ production, using this acceptance and
the theoretical cross section\cite{xsec}, is shown in Table~2.

The dilepton background from $WW$ production is calculated using ISAJET,
assuming a total $WW$ cross section of 9.5~pb\cite{ohnemus}, and is
found to be 0.16$\pm$0.06 events. $WW$ events may contain two jets due
to initial-state gluon radiation.
The treatment of initial-state radiation
in the ISAJET calculation is checked using $Z$+jets data, and good agreement
is found. The background from $Z\rightarrow\tau\tau$
is estimated using $Z\rightarrow ee$ data, where each electron is replaced
by a simulated $\tau$ that decays leptonically. This background contributes
0.13$\pm$0.04 events. 
We estimate backgrounds from $b\bar{b}$ and $c\bar{c}$ using 
ISAJET to model production processes, and the CLEO Monte Carlo 
program\cite{cleo} to model $B$-meson decay. The Monte Carlo rates
are normalized to a sample of $e\mu$ data collected with lower trigger
thresholds. We estimate $0.10\pm 0.06$ background events from these sources.
Backgrounds from hadrons misidentified as leptons (0.07$\pm$0.05 events)
and the Drell-Yan production of lepton pairs (0.10$^{+0.23}_{-0.08}$)
are estimated from inclusive-jet and $Z$ data respectively.
The total expected background is 0.56$^{+0.25}_{-0.13}$ events, with 
two candidates observed.
 

Events selected for the lepton+jets search are required to 
have an isolated lepton with $E_T$ ($P_T$ for muons) $>20$~GeV 
and $|\eta|<1.0$, and
to have $\met>20$~GeV\cite{jetcor}. Events containing $Z$
bosons are removed by rejecting events with an $ee$ or $\mu\mu$ invariant
mass between 70 and 110~GeV/c$^2$.
In Table~\ref{ljets} we classify the 
$W$ candidate events according to the multiplicity, $N_{jet}$,
of jets with $E_T>15$~GeV and $|\eta|<2.0$\cite{jetcor}.
The dominant background in the lepton+jets search is the direct 
production of $W$+jets. 
The ratio of the $\ttbar$ signal to $W+$jets background can be greatly
improved by requiring $N_{jet}\geq~3$. This requirement has a 
rejection factor of $\approx$400 against inclusive $W$ production while
keeping approximately 75\% of the $t\bar{t}$ signal in the lepton+jets
mode for $M_{top}=160$~GeV/c$^2$. In the $W+\geq$3-jet sample,
we expect 12$\pm$2 (6.6$\pm$0.7) $\ttbar$ events for $M_{top} = 160$ (180)
GeV/c$^2$, using the acceptance discussed below and the theoretical
cross section. We observe 52 events with $N_{jet}\geq 3$.

The VECBOS Monte Carlo program\cite{vecbos} can be used to make estimates
of direct $W$+jets production. Table~3 shows the results of a particular
calculation which predicts 46 events with $\geq$3 jets and seven
events with $\geq$4 jets.
The VECBOS predictions for $\geq$ 3 jets have uncertainties of about a factor
of two due to the choice of $Q^2$ scale and cannot be used for a reliable
absolute background calculation. We have therefore developed a technique for
estimating backgrounds in the lepton+jets search directly from the data.
This technique is described below.
Other backgrounds
(direct $b\bar{b}$, $Z$ bosons, $W$ pairs, and hadrons misidentified as 
leptons) contribute 12.2$\pm$3.1 events\cite{prd}. Additional background 
rejection is needed to isolate a possible $\ttbar$
signal. Requiring the presence of a $b$ quark, tagged either by
a secondary vertex or by a semileptonic decay, provides such rejection.


The lifetime of $b$ hadrons 
can cause the $b$-decay vertex to be
measurably displaced from the $\bar{p}p$ interaction vertex.
When associated with jets
with $E_T>15$~GeV and $|\eta|<2.0$,
SVX tracks with $P_T\geq$~2 GeV/c and 
impact-parameter significance $|d|/\sigma_{d}\geq$~3 are used in 
a vertex-finding algorithm\cite{prd}. Using these tracks, 
the decay length transverse to the beam, $L_{xy}$, and its
uncertainty (typically $\sigma_{L_{xy}}\approx 130~\mu$m) are calculated
using a three-dimensional fit, with the tracks constrained to
originate from a common vertex.
Jets that have a secondary vertex displaced in the 
direction of the jet, with
significance $|L_{xy}|/\sigma_{L_{xy}} \geq 3.0$, are 
defined to be ``SVX-tagged."

We use a control sample, enriched in $b$-decays, of inclusive electrons 
($E_T > 10$~GeV) to measure the efficiency for SVX-tagging a 
semileptonic $b$ jet. 
We compare this efficiency with that predicted by the ISAJET+CLEO
$\bbbar$ Monte Carlo and find our measured efficiency to be lower than
the Monte Carlo prediction by a factor of 
0.72$\pm$0.21. We then determine the efficiency for 
tagging at least one $b$ jet in a $\ttbar$ event with three or 
more observed jets, $\epsilon_{tag}$, from $\ttbar$ Monte Carlo rescaled 
by the factor determined above.
We find $\epsilon_{tag}=22\pm 6$\% independent of top
mass for $M_{top}>120$~GeV/c$^2$. The efficiency, $\epsilon_{SVX}$,
for inclusive $\ttbar$ events to pass the lepton-identification,
kinematic, and SVX $b$-tag requirements is shown in Table~1.
The number of expected SVX-tagged $\ttbar$ events
with $N_{jet}\geq 3$ is shown in Table~2. Six SVX-tagged events
are observed in the 52-event $W$+$\geq$3-jet sample.

Rather than rely on Monte Carlo predictions, we estimate directly from our 
data how many tags we would expect in the
52-event sample if it were entirely background. We assume
that the heavy-quark ($b$ and $c$) content of jets in $W$+jets background 
events is the same as in an inclusive-jet sample\cite{prd}. 
This assumption is expected to be conservative,
since the inclusive-jet sample contains heavy-quark contributions 
from direct production (e.g. $gg\rightarrow b\bar{b}$), 
gluon splitting (where a final-state
gluon branches into a heavy-quark pair), and flavor excitation
(where an initial-state gluon excites a heavy quark in the proton
or antiproton sea), while
heavy quarks in $W$+jets background events are expected to be produced
almost entirely from gluon splitting\cite{MLM}.
We apply the tag rates  measured in the inclusive-jet
sample, parametrized by the $E_T$ and track multiplicity of each jet, to the
jets in the 52 events to yield the total expected number of SVX-tagged events
from $Wb\bar{b}$, $Wc\bar{c}$, and fake tags due to track mismeasurement.
We have tested this technique in a number of control samples and use the
level of agreement with the number of observed tags to determine the
systematic uncertainty on the predicted tag rate.
The backgrounds from non-$W$ sources (direct $b\bar{b}$ production and
hadrons misidentified as leptons)
are also determined from the data\cite{prd}.
The small contributions from
$Wc$, from $WW$ and $WZ$ production, and from $Z\rightarrow\tau\tau$ 
are estimated
from Monte Carlo events. The total estimated background to SVX tags in 
the 52-event sample is 2.3$\pm$0.3 events. An alternate background estimate,
using Monte Carlo calculations of the heavy-quark processes in $W$+jets
events and a fake-tag estimate from jet data, predicts a heavy-quark
content per jet approximately a factor of three lower than in inclusive-jet
events and gives an overall background estimate a factor of 1.6 lower than
the number presented above, supporting the conservative nature of our 
background estimate.

In the $W$+jets sample, the $L_{xy}$ distribution of observed SVX tags 
is consistent with that of heavy-quark jets. The tags in the $W$ events 
with one and two jets are expected to come mainly from sources other 
than $\ttbar$ decay, and the
rate of these tags is consistent with the background prediction, with 16
events tagged and 22.1$\pm$4.0 predicted.

%

A second technique for tagging $b$ quarks is to search for leptons arising
from the decays $b\rightarrow\ell\nu X ~(\ell = 
e~{\rm or}~ \mu)$, or  $b\rightarrow c\rightarrow\ell\nu X$.  
Because these leptons typically have 
lower $P_T$ than leptons from $W$ decays, we refer to 
them as ``soft lepton tags", or SLT. We require lepton $P_T>2$~GeV/c.
To keep this analysis statistically independent of the dilepton search,
leptons that pass the dilepton requirements are not considered as SLT 
candidates.

In searching for electrons from $b$ and $c$ decays, each CTC track 
is extrapolated to the calorimeter, and a
match is sought
to an electromagnetic cluster consistent in size, shape, 
and position with expectations for electron showers. 
The efficiency of the electron selection criteria, excluding isolation cuts,
is determined from a sample of electron pairs
from photon conversions, where the first electron is identified in
the calorimeter and the second, unbiased, electron is selected using
a track-pairing algorithm. The electron isolation efficiency is
determined from $\ttbar$ Monte Carlo events. The total
efficiencies are (53$\pm$3)\% and (23$\pm$3)\% (statistical uncertainties
only) for electrons from $b$ and sequential $c$ decays respectively.
To identify muons, track segments in the muon chambers
are matched to tracks in the CTC. 
The efficiency for reconstructing track segments in the muon chambers is
measured to be 96\% using $J/\psi\rightarrow \mu^+\mu^-$ and 
$Z\rightarrow\mu^+\mu^-$ decays. This number is combined with the
$P_T$-dependent efficiency of the track-matching requirements to give an
overall efficiency of approximately 85\% for muons from both $b$ and
$c$ decays. 

The acceptance of the SLT analysis for $\ttbar$ events is 
calculated using the ISAJET and CLEO Monte Carlo programs. 
The efficiency for tagging at least one jet in a $\ttbar$ event by detecting
an additional lepton with $P_T>$ 2~GeV/c is
$\epsilon_{tag} = 16 \pm 2$\%, approximately independent of $M_{top}$.
The efficiency, $\epsilon_{SLT}$,
for inclusive $\ttbar$ events to pass the lepton-identification,
kinematic, and SLT $b$-tag requirements is shown in Table~1. The
number of expected SLT-tagged $\ttbar$ events is shown in Table~2.
We find seven SLT-tagged events with $N_{jet}\geq 3$. Three of the
seven also have SVX tags.

The main backgrounds to
the SLT search are hadrons misidentified as leptons,
and $Wb\bar{b}$, $Wc\bar{c}$ production. As in the SVX analysis,
we estimate these backgrounds from the data by conservatively assuming that
the heavy-quark content per jet in $W$+jets events is the same
as in inclusive-jet events. By studying tracks in such events, 
we measure the probability of misidentifying a hadron as an
electron or muon, or of tagging a true semileptonic decay. We use
these probabilities to predict the number of tags in a variety of 
control samples, and obtain good agreement with the number observed.
We expect 2.70$\pm$0.27 tags in the $W+\geq$3~jet sample
from these sources. Other sources (direct $\bbbar$, $W$/$Z$ pairs, 
$Z\rightarrow\tau\tau$, $Wc$, and Drell-Yan) contribute 0.36$\pm$0.09
events, for a total SLT background of 3.1$\pm$0.3 events.
The number of SLT tags in the $W+$1 and
$W$+2-jet samples, which should have only a small contribution from $\ttbar$,
agrees with the background expectation (45 events tagged, 44$\pm$3.4
predicted). 
Figure~\ref{combo-njet}
shows the combined number of SVX and SLT tags, together with the 
estimated background, as a function of jet multiplicity.


Each of the analyses presented above shows an excess of events over
expected backgrounds, as shown in Table~\ref{Table3}. The dilepton
analysis observes two events with a background of 0.56$^{+0.25}_{-0.13}$.
The lepton+jets $b$-tag analysis identifies ten events: six events with a 
background of 2.3$\pm$0.3 using the SVX tagging algorithm, and seven events 
with a background of 3.1$\pm$0.3 using the SLT tagging algorithm, with 
three of these events tagged by both algorithms.
For each of 
these results we calculate the probability, $\cal P$, that
the estimated background has fluctuated up to the number of candidate events
seen or greater. We find ${\cal P}_{DIL}$=12\%, ${\cal P}_{SVX}$=3.2\%,
and ${\cal P}_{SLT}$=3.8\%.

To calculate the probability ${\cal P}_{combined}$ that all three results
together are due only to an upward fluctuation of the background, 
we use the observation of 15 ``counts": the two dilepton events, 
the six SVX tags, and the seven SLT tags.
This procedure gives extra weight to the double-tagged events, which are
approximately six times more likely to come from $b$ and $c$ jets
than from fakes, and therefore have a significantly smaller background
than the single-tagged events.  We have checked that we understand SVX$-$SLT
correlations by correctly predicting the number of
double-tagged jets and events in the inclusive-jet sample.
We calculate ${\cal P}_{combined}$ using a Monte Carlo
program that generates many samples of 52 background events, with
fractions of $W$+light quark and gluon jets, $Wb\bar{b}$, $Wc\bar{c}$, and 
other backgrounds distributed according to Poisson statistics with 
mean values and uncertainties
predicted by Monte Carlo calculations\cite{prd}. The number of events with
heavy-quark jets is scaled up to agree with the more conservative background 
estimate from inclusive-jet data.
The predicted number of 
SVX plus SLT-tagged events is obtained by applying the measured efficiencies 
and correlations in the SVX and SLT fake rates. This number is combined
with a Poisson-distributed number of dilepton background events to
determine the fraction of experiments with 15 or more counts from
background alone. We find 
${\cal P}_{combined}$=0.26\%. This corresponds to a 2.8$\sigma$ excess for
a Gaussian probability function.

Assuming the excess events to be from $\ttbar$, we calculate the cross
section for $\ttbar$ production in $\ppbar$ collisions at $\sqrt{s}$=1.8~TeV.
The calculation uses the $\ttbar$ acceptance,
the derived efficiencies for tagging jets in $\ttbar$ events and
a revised estimate of the background appropriate for a mixture of
$\ttbar$ events and background in the 52-event $W$+jets sample
(rather than assuming it to contain all background as above).
In Tables~\ref{sum_acc} and~\ref{xsec-table} we summarize the acceptances, 
and the theoretical and measured cross sections as a function of $M_{top}$.

We have also studied\cite{prd} distributions of kinematic
quantities for the 52 $W$+$\geq$3~jet events. If the top quark is 
very massive the decay jets
will typically be more energetic than jets in $W$+jets background events.
One variable with discrimination is $E_{T23}=(E_{T2}+E_{T3})$, where
$E_{T2}$ and $E_{T3}$ refer to the $E_T$'s of the second- and third-most
energetic jets in the event. The VECBOS Monte Carlo program
predicts that in $W+$jets background events the median of 
$E_{T23}$ is 71~GeV, while 93\% of HERWIG\cite{monte_carlos}
$\ttbar$ events ($M_{top}=160$~GeV/c$^2$) have $E_{T23}>71$~GeV. 
In the 52-event sample, 39 events have $E_{T23}>71$~GeV, as do eight
of the ten $b$-tagged events. This is qualitatively consistent with the 
$\ttbar$ hypothesis; however additional studies in progress 
are needed to reduce systematic uncertainties on the 
jet energy scale and on the $E_{T23}$ distribution of the background.


Assuming that the excess of $b$-tagged events is due to $t\bar{t}$ production,
we estimate $M_{top}$ using a constrained fit\cite{SQUAW} 
to each tagged event with four~jets. Using the 52-event $W$+$\geq$3-jet sample,
we require a fourth jet with $\Et > 8$~GeV and $|\eta| <$~2.4.
Seven of the ten $b$-tagged events identified in the 
lepton+jets analysis pass this requirement. These seven 
events are fitted individually to the hypothesis
that three of the jets come from one $t$ or $\bar{t}$ through its decay
to $Wb$, and that the lepton, $\met$, and the remaining jet come from the other
$t$ or $\bar{t}$ decay\cite{jetcor}. If the event contains 
additional jets, only the four highest-$E_T$ jets are used in the fit.
The fit is made for all six jet configurations, with the requirement 
that the tagged jet in the event must be one of the $b$ quarks. 
There are two solutions in each case for the longitudinal momentum of 
the neutrino, and the one corresponding to the best $\chi^2$ is chosen.

Application of this method to $\ttbar$ Monte Carlo events
($M_{top} = 170$~GeV/c$^2$)
gives a distribution with a peak at 168~GeV/c$^2$
and a rms spread of 23~GeV/c$^2$. 
Fitting Monte Carlo $W+$jets background events to 
the $\ttbar$ hypothesis yields a
mass distribution with a broad peak centered at about 140 GeV/c$^2$.

The results of the fits to the seven events are presented in 
Figure~\ref{mass_fig}. In this sample, 1.4$_{-1.1}^{+2.0}$ events are 
expected to come from background\cite{prd}.
To find the most likely top mass from the seven events,
we perform a likelihood fit of their mass distribution to 
a sum of the expected distributions from
$W$+jets and a top quark of mass $M_{top}$. The \mbox{$-\log$(likelihood)}
distribution from this fit is shown in the inset to Figure~2.
Systematic uncertainties in this fit arise from the background estimation,
the effects of gluon radiation on the determination of parton energies,
the jet energy scale, kinematic bias in the tagging algorithms, 
and different methods of performing the likelihood fit.
Combining these uncertainties yields a top mass of 
$M_{top} = 174 \pm 10 _{-12}^{+13}$ GeV/c$^2$, where the first uncertainty is 
statistical and the second is systematic. The statistical uncertainty
includes the effects of detector resolution and incorrect assignments of 
jets to their parent partons. Using the acceptance
for this top mass and our measured excess over background we find
$\sigma_{\ttbar}(M_{top}=174~\rm{GeV}/c^2) = 13.9^{+6.1}_{-4.8}$~pb.
By performing a simple $\chi^2$ analysis on 
the theoretical prediction for the cross section as
a function of $M_{top}$, our measured mass, and our measured
cross section, we find that the three results are compatible
at a confidence level of 11\% (1.6$\sigma$).


We have performed many consistency checks, and have found
some features of the data that do not support the $\ttbar$ hypothesis.
The sample of inclusive $Z$ events serves as a control sample for studying
the production of a vector boson plus jets, as $Z$ bosons are not 
produced in $\ttbar$ decay.
We find two $b$-tagged $Z$+$\geq$3~jet events with 0.64 expected. 
Both events have four jets and are SVX-tagged. Though the statistics are 
limited, these events could indicate an 
additional (non-$\ttbar$) source of vector boson plus heavy quark production, 
not accounted for in our background estimates.
Higher-statistics
checks of the $b$-tagging rate in $W$ or $Z$+1 and 2-jet events are consistent
with expectations. 
We also find that the measured $\ttbar$ cross section
is large enough to account for all observed $W$+4~jet events.
The apparent deficit of events from direct production of $W$+4~jets and
other backgrounds is a 1.5-2$\sigma$ effect.

Other features do support the $\ttbar$ hypothesis. One of the
dilepton candidate events is $b$-tagged by both the SVX and SLT algorithms,
with approximately 0.01 double-tagged background events (0.13 signal events) 
expected. This, together with the excess of $b$-tagged $W$+jets events, provides
evidence for an excess of both $Wb\bar{b}$ and $WWb\bar{b}$ production,
as expected from $\ttbar$ decays. We have performed a kinematic analysis of
the lepton+jets sample and conclude that it can accommodate
the top content implied by our measured cross section. 
Furthermore, a likelihood fit to the top mass distributions obtained from
the $b$-tagged $W$+4-jet events prefers the $\ttbar$+background
hypothesis over the background-only hypothesis by 2.3 standard deviations.

In conclusion, the data presented here give evidence for, but do not
firmly establish, the existence of the top quark. Work is continuing
on kinematic analyses of the present data, and we hope for an 
approximate four-fold increase in data from the 1994-95 Tevatron 
collider run.

This work would not have been possible without the skill and                   
hard work of the Fermilab staff. We thank                 
the staffs of our institutions for their many contributions                    
to the construction of the detector.  We also thank Walter Giele for
advice and many helpful suggestions regarding $W$+jets and the VECBOS
Monte Carlo program, and Gerry Lynch for help with the kinematic
fitting program.
This work is supported by the U.S. Department of                              
Energy; the National Science Foundation;
the Italian Istituto Nazionale di Fisica Nucleare;
the Ministry of Education, Science and Culture of Japan;
the Natural Sciences and Engineering Council of Canada;
the National Science Council of the Republic of China;
the A.P. Sloan Foundation; and the Alexander von Humboldt-Stiftung.
\vskip 0.5in

$^{(a)}$ Visitor.

\newpage
\begin{samepage}

\end{samepage}

\newpage
\begin{table}[h]
\begin{center}
\vspace{0.25in}
\begin{tabular}{lcccc}
\hline \hline
 $M_{top}$ &  120 GeV/c$^2$ &
140 GeV/c$^2$ & 160 GeV/c$^2$ & 180 GeV/c$^2$\\ 
\hline 
 $\epsilon_{DIL}$ & $0.49 \pm .07\%$ & $0.66 \pm .07\%$ & 
                            $0.78 \pm .07\%$ & $0.86 \pm .07\%$           \\
 $\epsilon_{SVX}$ & $1.0 \pm 0.3\%$ & $1.5 \pm 0.4\%$ & $1.7 \pm 0.5\%$ & 
                           $1.8 \pm 0.6\%$                                  \\
 $\epsilon_{SLT}$ & $0.84 \pm 0.17 \%$ & $1.1 \pm 0.2\%$ & $1.2 \pm 0.2 \%$
                            & $1.3 \pm 0.2 \%$ \\
\hline
 $\sigma_{\ttbar}^{Theor}$ (pb) 
                   & $38.9^{+10.8}_{-5.2}$ & $16.9^{+3.6}_{-1.8}$ 
                   & $8.2^{+1.4}_{-0.8}$   & $4.2^{+0.6}_{-0.4}$     \\
\hline
 $\sigma_{\ttbar}^{Expt}$(pb) & $22.7^{+10.0}_{-7.9}$ & $16.8^{+7.4}_{-5.9}$
                            & $14.7^{+6.5}_{-5.1}$ & $13.7^{+6.0}_{-4.7}$  \\
\hline \hline
\end{tabular}
\end{center}
\caption{Summary of top acceptance (including branching ratios) 
and the theoretical cross section\protect\cite{xsec}. The last line gives the 
$\ttbar$ production cross section obtained from this measurement.}
\label{Table2}
\label{sum_acc}
\end{table}
\begin{table}[h]
\begin{center}
\vspace{0.25in}
\begin{tabular}{lccc}
\hline \hline
 Channel: &  Dilepton & SVX & SLT \\ \hline 
 $N_{expected}$, $M_{top}=120$ GeV/c$^{2}$ & $3.7 \pm 0.6$ & $7.7 \pm 2.5$ &
		$6.3 \pm 1.3$  \\
 $N_{expected}$, $M_{top}=140$ GeV/c$^{2}$  & $2.2 \pm 0.2$ & $4.8 \pm 1.7$ &
		$3.5 \pm 0.7$ \\
 $N_{expected}$, $M_{top}=160$ GeV/c$^{2}$ & $1.3 \pm 0.1$ & $2.7 \pm 0.9$ &
		$1.9 \pm 0.3$  \\
 $N_{expected}$, $M_{top}=180$ GeV/c$^{2}$ & $0.68 \pm 0.06$ & $1.4 \pm 0.4$ &
		$1.1 \pm 0.2$ \\
\hline
 Total Background & $0.56^{+0.25}_{-0.13}$ & $2.3 \pm 0.3$ & $3.1 \pm 0.3$ \\
\hline
 Observed Events   & $2$ & $6$ & $7$ \\ \hline \hline
\end{tabular}
\end{center}
\caption{Number of $\ttbar$ events expected assuming the theoretical
cross section, and the number of candidate 
events observed with 
expected backgrounds.} 
\label{Table3}
\label{xsec-table}
\end{table}
\clearpage
%
%
\begin{table}[h]
\begin{center}
\vspace{0.25in}
\begin{tabular}{lcccc}
\hline\hline
$N_{jet}$ & Electrons & Muons & 
Total & VECBOS ($Q^{2}$=$<{P_T}^2$\mbox{}$>$)\\
\hline      
   0 Jet       & 10,663 & 6,264 & 16,927 & ------ \\
   1 Jet       &   1058 &   655 &   1713 & $1571^{+285}_{-227}$ \\
   2 Jets      &    191 &    90 &    281 & $ 267^{+80}_{-57}$ \\
   3 Jets      &     30 &    13 &     43 & $  39^{+12}_{-10}$   \\
 $\geq$~4 Jets &      7 &     2 &      9 & $   7^{+3.2}_{-2.2}$ \\
\hline\hline
\end{tabular}
\end{center}
\caption{Summary of $W$ candidate event yields as a function of 
jet multiplicity.
Jets have $\Et \ge$ 15 GeV and $|\eta| \le$ 2.0. Also shown are the
predicted number of $W$ events from the VECBOS Monte Carlo program.
The uncertainties shown in the VECBOS predictions are dominated by the 
uncertainty in the jet energy scale; the uncertainty in the 
$Q^2$-scale is not included.}
\label{Table1}
\label{ljets}
\end{table}
\clearpage

\newpage
\begin{figure}[h]
\vskip 6in
\caption{The sum of SVX and SLT tags observed in the $W+$jets data
 (solid triangles).
Events tagged by both algorithms are counted twice.  The shaded area is
the sum of the background estimates for SVX and SLT, with its uncertainty.
The three-jet and four-or-more-jet bins are the $\ttbar$ signal region.}
\label{combo-njet}
\end{figure}

\newpage
\begin{figure}[h]
\vskip 6in
\caption{Top mass distribution for the data (solid histogram),
the $W$+jets background (dots), and the sum of 
background + Monte Carlo $\ttbar$ for $M_{top} = 175$~GeV/c$^2$ (dashed).
The background distribution has been normalized to the 1.4 
background events expected in the mass-fit sample. The inset shows
the likelihood fit used to determine the top mass.}
\label{mass_fig}
\end{figure}


\begin{thebibliography}{99}
\bibitem{coord} Here $P_T = P\sin\theta$, where $\theta$
is the polar angle with respect to the proton beam direction. The
pseudorapidity, $\eta$, is defined as $-\ln\tan\frac{\theta}{2}$.
\bibitem{jetcor} Corrections to the observed jet energy and $\met$ are
   described in Ref.~\cite{prd} and references therein.
\end{thebibliography}
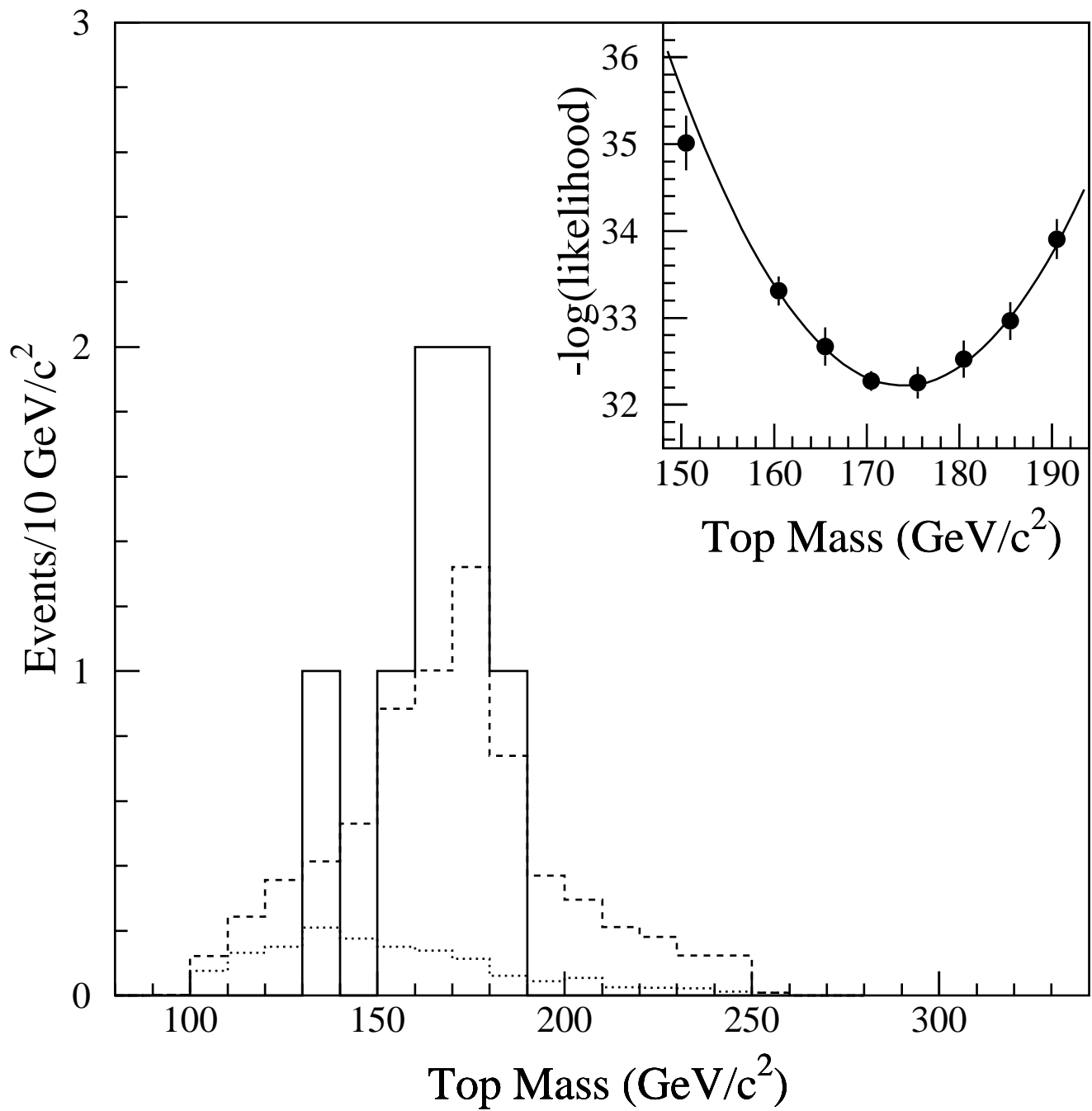
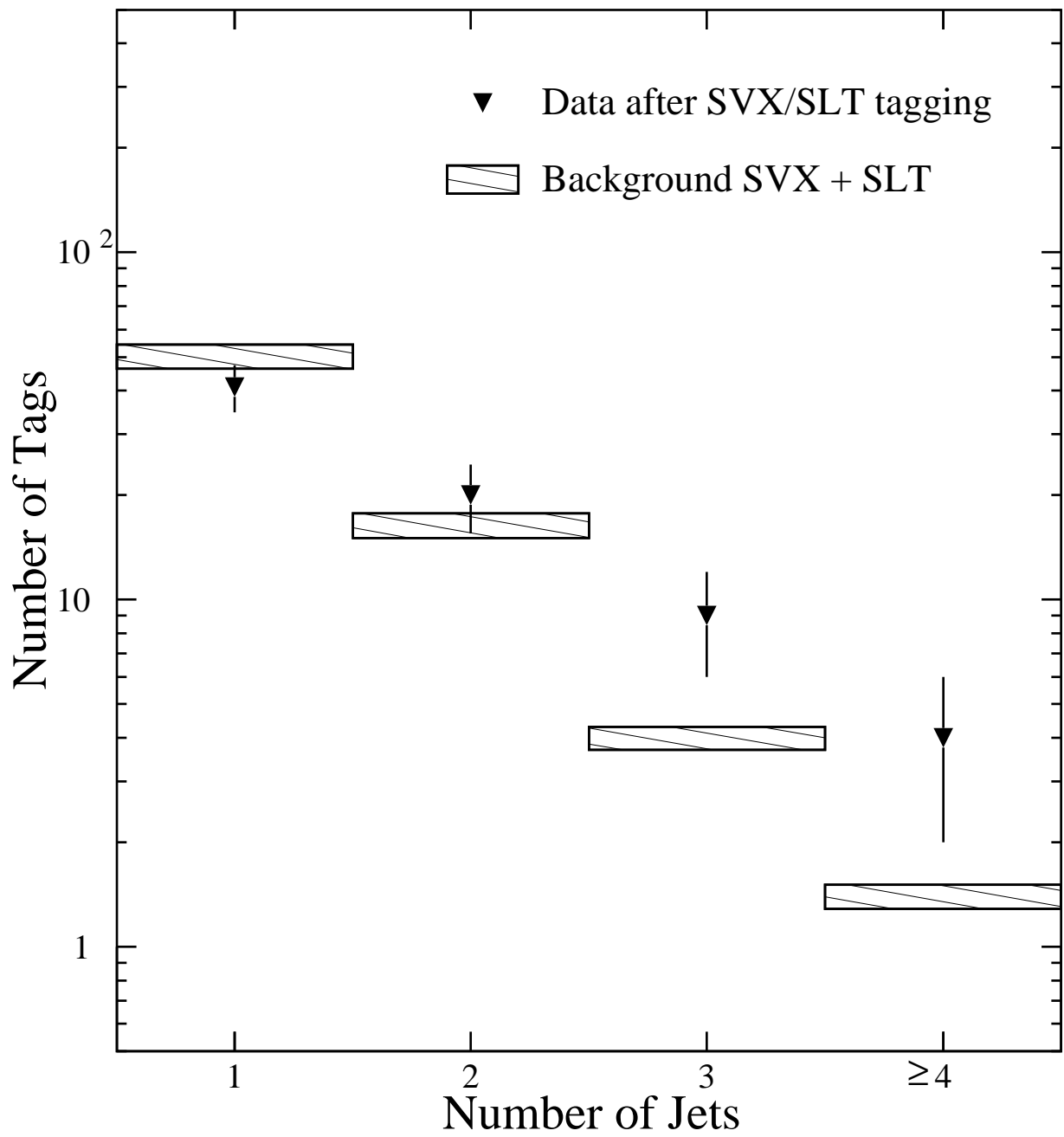
\end{document}